\documentclass[prd,twocolumn,showpacs,superscriptaddress]{revtex4-1}
\usepackage{graphicx}
\usepackage{units}
\usepackage{multirow}
\usepackage{bigstrut}
\usepackage{overpic}
\usepackage{array}
\usepackage{color}
\usepackage{amsmath}
\usepackage{bm}
\usepackage{times}
\RequirePackage{lineno}
\usepackage[bookmarksnumbered, pdfstartview=FitH,colorlinks,urlcolor=blue, citecolor=blue,linkcolor=blue] {hyperref}
\begin{document}

\modulolinenumbers[1]

\setlength{\oddsidemargin}{-0.5cm} \addtolength{\topmargin}{15mm}

\title{\boldmath Study of the decay $D^0\rightarrow \rho(770)^-e^+\nu_e$ }

\author{
M.~Ablikim$^{1}$, M.~N.~Achasov$^{4,c}$, P.~Adlarson$^{76}$, O.~Afedulidis$^{3}$, X.~C.~Ai$^{81}$, R.~Aliberti$^{35}$, A.~Amoroso$^{75A,75C}$, Q.~An$^{72,58,a}$, Y.~Bai$^{57}$, O.~Bakina$^{36}$
% ...
\\
\vspace{0.2cm}
(BESIII Collaboration)\\
\vspace{0.2cm} {\it
$^{1}$ Institute of High Energy Physics, Beijing 100049, People's Republic of China\\
% ...
$^{3}$ Bochum  Ruhr-University, D-44780 Bochum, Germany\\
$^{4}$ Budker Institute of Nuclear Physics SB RAS (BINP), Novosibirsk 630090, Russia\\
% ...
$^{35}$ Johannes Gutenberg University of Mainz, Johann-Joachim-Becher-Weg 45, D-55099 Mainz, Germany\\
$^{36}$ Joint Institute for Nuclear Research, 141980 Dubna, Moscow region, Russia\\
% ...
$^{57}$ Southeast University, Nanjing 211100, People's Republic of China\\
$^{58}$ State Key Laboratory of Particle Detection and Electronics, Beijing 100049, Hefei 230026, People's Republic of China\\
% ...
$^{72}$ University of Science and Technology of China, Hefei 230026, People's Republic of China\\
% ...
$^{75}$ University of Turin and INFN, (A)University of Turin, I-10125, Turin, Italy; (B)University of Eastern Piedmont, I-15121, Alessandria, Italy; (C)INFN, I-10125, Turin, Italy\\
$^{76}$ Uppsala University, Box 516, SE-75120 Uppsala, Sweden\\
% ...
$^{81}$ Zhengzhou University, Zhengzhou 450001, People's Republic of China\\
\vspace{0.2cm}
$^{a}$ Deceased\\
% ...
$^{c}$ Also at the Novosibirsk State University, Novosibirsk, 630090, Russia\\
% ...
}
}

\begin{abstract}
We present a   study of the semileptonic decay  $D^0\rightarrow \pi^-\pi^0e^{+}\nu_{e}$  using an $e^+e^-$ annihilation data sample of  $7.93~\mathrm{fb}^{-1}$ collected at the center-of-mass energy of 3.773 GeV with the BESIII detector.
The branching fraction of $D^0\to \rho(770)^-e^+\nu_e$ is measured  to be  $(1.439 \pm 0.033(\rm stat.) \pm 0.027(\rm syst.)) \times10^{-3}$, which is a factor 1.6 more precise than previous measurements.  By performing an amplitude analysis,  we measure the hadronic form-factor ratios of $D^0\to \rho(770)^-e^+\nu_e$  at $q^2=0$  assuming the single-pole-dominance parametrization: $r_{V}=V(0)/A_1(0)=1.548\pm0.079(\rm stat.)\pm0.041(\rm syst.)$ and $r_{2}=A_2(0)/A_1(0)=0.823\pm0.056(\rm stat.)\pm0.026(\rm syst.)$. 
\end{abstract}

\maketitle

\section{Introduction}
Studies of semileptonic\,(SL) decays of charm mesons provide an ideal test-bed to explore the weak and strong interactions in mesons composed of heavy quarks~\cite{pr494_197,ARNPS73_285-314}.  The SL partial-decay width is related to the product of the hadronic form factors, which describe the strong interactions between final-state quarks, including non-perturbative effects, and the Cabibbo-Kobayashi-Maskawa  matrix elements~\cite{prl10_531, ptp49_652}. Thus, precise measurements of form factors are important to test different theoretical models and improve the inputs of theoretical calculations~\cite{prd52_2783,prd44_3567}.
Recently, various theoretical models have been used to calculate the hadronic transition form factors in $D^0\rightarrow \rho(770)^- e^+\nu_e$,  including   the covariant quark model (CQM), the covariant confining-quark model (CCQM),  the light-front quark model (LFQM),   light-cone sum rule (LCSR) calculations~\cite{cqm_2000,fpb14_66401,lfqm_2012,Ijmp21_6125-6172}, and  the HM$\chi$T model  (based on the combination of heavy meson and chiral symmetries)~\cite{hmt_2005}.  Their theoretical calculation results are summarized in Table~\ref{tab:Theory}. The wide range in the predictions of the ratio $r_2$, in particular, offer an opportunity for measurement to distinguish between these models.

\begin{table}[tp!]
\caption{
The theoretical calculation results of the hadronic form-factor ratios $ r_V$ and $r_2$ for $D^0\rightarrow \rho(770)^- e^+\nu_e$.}
\begin{center}
\begin{tabular}
{lccc} \hline\hline Theory        & $ r_V$     &  $r_2$     \\
       	\hline
                CQM~\cite{cqm_2000}  &   1.53 &  0.83\\
                CCQM~\cite{ fpb14_66401}  &  1.26  $\pm $   0.25 &  0.93  $ \pm$ 0.19\\
		LFQM~\cite{lfqm_2012}  & 1.47  &   0.78\\	
		LCSR~\cite{Ijmp21_6125-6172} & 1.34  &  0.62\\
	        HM$_\chi$T~\cite{hmt_2005}     & 1.72  &  0.51 \\
\hline\hline
\end{tabular}
\label{tab:Theory}
\end{center}
\end{table}

In this paper, an improved measurement of the  branching fraction (BF) for $D^0\rightarrow \rho(770)^- e^+\nu_e$  is presented, along with an analysis of the decay dynamics. These measurements are performed using an $e^+e^-$ annihilation data sample corresponding to an integrated luminosity of
$7.93~\mathrm{fb}^{-1}$ collected  at $\sqrt{s}=3.773$ GeV  with the BESIII detector~\cite{Ablikim:2009aa,Ablikim:2024lum,Ablikim:2023} at the BEPCII collider.   Charge-conjugate modes are implied throughout this paper.

\section{BESIII Detector and Monte Carlo Simulation}
The BESIII detector  records symmetric $e^+e^-$ collisions  provided by the BEPCII storage ring in the center-of-mass energy range from 2.0 to 4.95~GeV,
with a peak luminosity of $1.1 \times 10^{33}\;\text{cm}^{-2}\text{s}^{-1}$  achieved at $\sqrt{s} = 3.773\;\text{GeV}$~\cite{Ablikim:2009aa,Future_040001}. The cylindrical core of the BESIII detector covers 93\% of the full solid angle and consists of a helium-based multilayer drift chamber~(MDC), a plastic scintillator time-of-flight system~(TOF), and a CsI(Tl) electromagnetic calorimeter~(EMC), which are all enclosed in a superconducting solenoidal magnet providing a 1.0~T magnetic field~\cite{detector,Future_040003}. The charged-particle momentum resolution at $1~{\rm GeV}/c$ is $0.5\%$, and the  ${\rm d}E/{\rm d}x$ resolution is $6\%$ for electrons from Bhabha scattering~\cite{Ablikim:2009aa}. The EMC measures photon energies with a resolution of $2.5\%$ ($5\%$) at $1$~GeV in the barrel (end-cap) region.  The time resolution in the TOF barrel region is 68~ps, while that in the end-cap region was 110~ps.  The end-cap TOF system was upgraded in 2015 using multigap resistive plate chamber technology, providing a time resolution of 60~ps, which benefits 63\% of the data used in this analysis~\cite{etof1,etof2,etof3}.

Simulated data samples produced with a {\sc geant4}-based~\cite{geant4} Monte Carlo (MC) package, which includes the geometric description of the detector and the detector response, are used to determine signal detection efficiencies and to estimate potential backgrounds. The simulation models the beam-energy spread and initial-state radiation (ISR) in the $e^+e^-$ annihilations with the generator {\sc kkmc}~\cite{kkmc}. The inclusive MC sample includes the production of $D\bar{D}$ pairs (including quantum coherence for the neutral $D$ channels), the non-$D\bar{D}$ decays of the $\psi(3770)$, the ISR production of the $J/\psi$ and $\psi(3686)$ states, and the continuum processes. All particle decays are modelled with {\sc evtgen}~\cite{evtgen} using BFs either taken from the
Particle Data Group~\cite{pdg16}, when available, or otherwise estimated with {\sc lundcharm}~\cite{lundcharm}. Final-state radiation from charged final-state particles is incorporated using the {\sc photos} package~\cite{photos}. The form factors obtained in this work are used to generate the signal MC sample of $D^0\rightarrow \rho^{-} e^{+}\nu_{e}$.

\section{Analysis}
At $\sqrt{s}=3.773$ GeV, the $D^0$ and $\bar{D}^0$ mesons are produced in pairs via the $e^+e^- \to \psi(3770) \to D^0\bar{D}^0$ process.
Events with $\bar{D}^0$ meson candidates fully reconstructed in one of the final states listed in  Table~\ref{tab:numST}, are denoted as single-tag (ST) events. In ST events, the $D^0 \to \pi^- \pi^0 e^+ \nu_e$ candidates are reconstructed in the recoil side of the $\bar{D}^0$ meson to form double-tag (DT) events.
The BF of $D^0\rightarrow \pi^-\pi^0e^{+}\nu_{e}$ is given by
\begin{equation}
\label{eq:branch}
{\mathcal B}_{\rm SL}=N_{\mathrm{DT}}/(N_{\mathrm{ST}}^{\rm tot}\cdot \epsilon_{\rm SL}\cdot{\mathcal B}_{\pi^0 \to \gamma \gamma }),
\end{equation}
where $N_{\rm ST}^{\rm tot}$ and $N_{\rm DT}$  are the yields of the ST and DT candidates in data, respectively, and ${\mathcal B}_{\pi^0 \to \gamma \gamma }$ is the BF of the $\pi^0 \to \gamma \gamma$ decay. The factor $\epsilon_{\rm SL}=\Sigma_i [(\epsilon^i_{\rm DT}\cdot N^i_{\rm ST})/(\epsilon^i_{\rm ST}\cdot N^{\rm tot}_{\rm ST})]$ is the efficiency of detecting the SL decay in the presence of the ST $\bar{D}^0$ meson. 
Here $i$ represents the tag mode, $N^i_{\rm ST}$ is the ST yield of tag mode $i$, $\epsilon^i_{\rm ST}$ is the ST efficiency of tag mode $i$, and $\epsilon^i_{\rm DT}$ is the DT efficiency of tag mode $i$. $N^i_{\rm ST}$ and $\epsilon^i_{\rm ST}$ are obtained from the data and inclusive MC sample, respectively, while $\epsilon^i_{\rm DT}$ is determined with the signal MC sample.

Charged tracks detected in the MDC are required to be within a polar angle ($\theta$) range of $|\! \cos\theta|<0.93$, where $\theta$ is the angle between the direction of charged track and the symmetry axis of the MDC, $z$-axis.
For  charged tracks not originating from $K_S^0$  decay, the distance of closest approach to the interaction point (IP) 
must be less than 10\,cm along the $z$-axis, $|V_{z}|$,   and less than 1\,cm in the transverse plane, $|V_{xy}|$.
Particle identification~(PID) of charged kaons and pions is performed using the d$E$/d$x$  measured by the  MDC and the flight time in the TOF. For the PID of the positron, extra EMC information is used to construct likelihoods for the positron, pion and kaon hypotheses ($\mathcal{L}_e$, $\mathcal{L}_\pi$ and $\mathcal{L}_K$).  The pion  and kaon  candidates are  required to satisfy  $\mathcal{L}_\pi > \mathcal{L}_K$ and  $\mathcal{L}_K > \mathcal{L}_\pi$, respectively, while the positron  candidate must satisfy $\mathcal{L}_e/(\mathcal{L}_e+\mathcal{L}_\pi+ \mathcal{L}_K) > 0.8$ and $\mathcal{L}_{e} > 0.001$. Additionally, an extra requirement of $E / (c \cdot p_{e^+}) > 0.8$ for those positron candidates is applied, where $E$ is the energy deposited in the EMC and $p_{e^+}$ is the momentum of the positron measured by the MDC. 

The $K_{S}^0$ candidate is reconstructed from two oppositely charged tracks, assumed to be $\pi^+\pi^-$ without imposing PID selection criteria, satisfying $|V_{z}|<$ 20~cm. They are constrained to originate from a common vertex and are required to have an invariant mass within $|M_{\pi^{+}\pi^{-}} - m_{K_{S}^{0}}|<$ 12~MeV$/c^{2}$, where $m_{K_{S}^{0}}$ is the  known $K^0_{S}$  mass~\cite{pdg16}. 
The decay length of the $K^0_S$ candidate, defined as the flight distance between the common vertex and the IP, is required to be greater than twice its  resolution. The quality of the vertex fits (primary-vertex fit and secondary-vertex fit) is ensured by a requirement  on the $\chi^2$ ($\chi^2<$100).
  The $\pi^0$ candidates are reconstructed  via the  $\pi^0\rightarrow \gamma\gamma$ decay, with photon candidate identified using showers in the EMC. While, the $\pi^0$  candidates with both photons from the end cap of the EMC are rejected due to poor resolution.
The deposited energy of each shower must be more than 25~MeV in the barrel region ($|\!\cos \theta|< 0.80$) or more than 50~MeV in the end-cap region ($0.86 <|\!\cos \theta|< 0.92$).
The opening angle between the photon candidate and the nearest charged track is required to be greater than $10^{\circ}$. The difference in EMC time from the event-start time is required to be within [0, 700] ns. 
  For any $\pi^0$ candidate, the invariant mass of the candidate photon pair must be within $(0.115,~0.150)$~GeV$/c^2$. To improve the momentum resolution, a one-constraint (1-C) kinematic fit is performed to constrain the $\gamma\gamma$ invariant mass to the known $\pi^0$ mass~\cite{pdg16}. The $\chi^2$ of the 1-C kinematic fit is required to be less than 50.  

To  identify the ST $\bar{D}^0$ mesons, we define two variables: the energy difference $\Delta E\equiv E_{\bar D^0}-E_{\mathrm{beam}}$ and the beam-constrained mass $M_{\rm BC}\equiv\sqrt{E_{\mathrm{beam}}^{2}/c^4-|\vec{p}_{\bar D}|^{2}/c^2}$, where $E_{\mathrm{beam}}$ is the beam energy, and $E_{\bar D^0}$ and $\vec{p}_{\bar D^0}$ are the total energy and momentum of the ST $\bar{D}^0$ meson in the $e^+e^-$ center-of-mass frame. If multiple $\bar{D}^0$ candidates are present in a given ST mode, the one with the minimum $|\Delta E|$ is kept for further analyses.  To suppress the combinatorial background in the $M_{\rm BC}$ distribution, $\Delta E$ requirements are imposed on the ST candidate events for each ST mode.
The requirements for $\Delta E$ and the ST efficiencies are decided upon by analyzing the inclusive MC sample and are summarized in Table~\ref{tab:numST}.

\begin{figure*}[tp!]
\begin{center}
\includegraphics[width=0.75\linewidth]{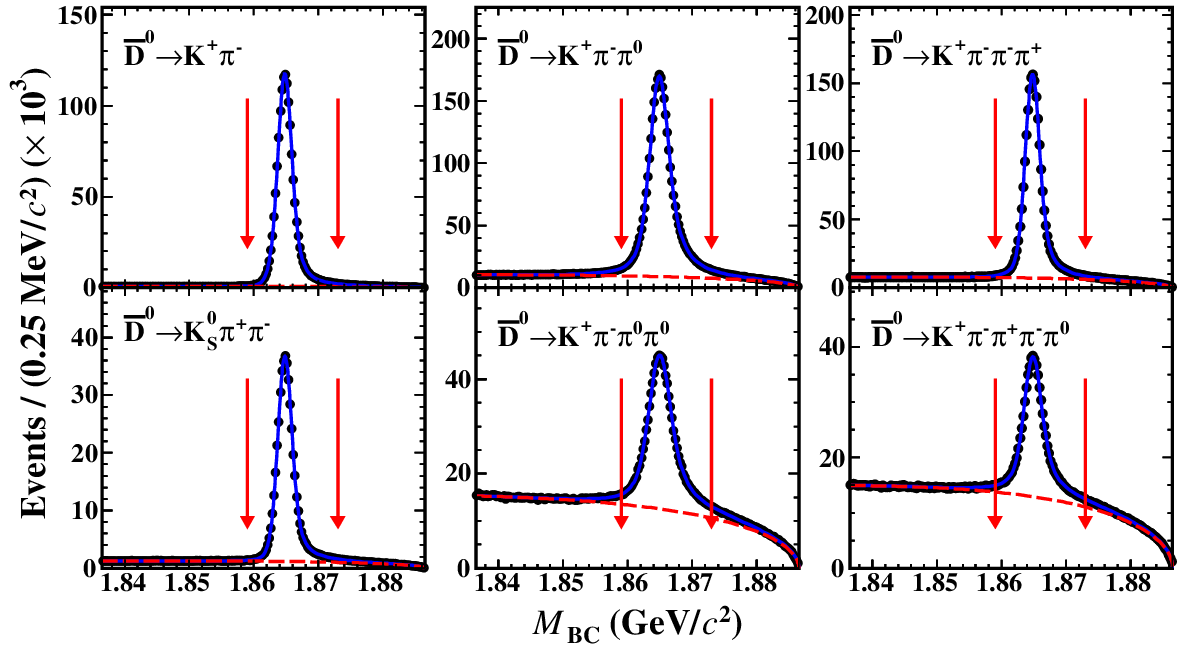}
\caption{ Fits to the $M_{\rm BC}$ distributions of the ST $\bar{D}^0$ candidates. The points with error bars  are data,
the blue curves are the best fits, and the red dashed curves are the fitted combinatorial background shapes.  The red
arrows indicate the $M_{\rm BC}$ signal window.}
\label{fig:tag_md0}
\end{center}
\end{figure*}

In each ST mode, the yield of ST $\bar{D}^0$ mesons is determined by fitting the corresponding $M_{\rm BC}$ distribution. The signal shape in the fit is obtained by  the MC-simulated signal shape  convolved with a double-Gaussian function to describe the resolution difference between MC simulation and data.
The combinatorial background shape is described using an ARGUS function~\cite{plb241_278}, with the endpoint fixed at 1.8865~GeV/$c^2$ corresponding to $E_{\mathrm{beam}}$. Figure~\ref{fig:tag_md0} shows the fits to the $M_{\rm BC}$ distributions of the accepted ST candidates in data for different ST modes.  Candidates with $M_{\rm BC}$ falling within the range of (1.859, 1.873) GeV/$c^2$  are retained for further analysis.
The yields for each ST mode are listed in Table~\ref{tab:numST}. Summing over the tag modes gives the total yield of ST $\bar{D}^0$ mesons to  be $N^{\rm tot}_{\rm ST}=(7895.8\pm3.4({\rm stat.}))\times10^{3}$.

\begin{table}[tp!]
\caption{ The $\Delta E$ requirements, the ST efficiencies ($ \epsilon^{i}_{\rm ST}$),  the obtained ST $\bar{D}^0$ yields in the data ($N^{i}_{\rm ST}$), and the DT efficiencies ($ \epsilon^{i}_{\rm DT}$).  The efficiencies do not include the BFs for  $K_{S}^0 \to \pi^+\pi^- $ and  $\pi^0 \to \gamma\gamma$. The uncertainties are statistical only.}
\begin{center}
\scalebox{0.85}{
\begin{tabular}
{lccccr} \hline\hline ST mode      & $\Delta E$ (GeV)        & $ \epsilon^{i}_{\rm ST}~(\%)$     &  $N^{i}_{\rm ST}$ ($\times 10^3$)    & $ \epsilon^{i}_{\rm DT}~(\%)$   \\
       	\hline
		$ K^{+}\pi^{-}$           & $(-0.027, 0.027)$ & 65.34  $\pm $   0.01  & 1449.5  $\pm $   1.3  & 20.67 $\pm$ 0.04\\
		$K^{+}\pi^{-}\pi^{0}$    &  $(-0.062, 0.049) $  & 35.59  $\pm $   0.01  & 2913.2  $\pm $  2.0 & 10.74 $\pm $ 0.02 \\
		$ K^{+}\pi^{+}\pi^{-}\pi^{-}$  & $(-0.026, 0.024)$ &  40.83  $\pm $   0.01 &  1944.2  $ \pm$ 1.6 & 11.69 $\pm $ 0.02\\
		$ K_S^{0}\pi^{+}\pi^{-}$  &  $(-0.024, 0.024)$ & 37.49  $\pm $   0.01  &   ~447.7  $ \pm$ 0.7 & 11.15 $\pm $ 0.04\\
		$K^{+}\pi^{-}\pi^{0}\pi^{0}$  &  $(-0.068, 0.053)$ & 14.83  $\pm $   0.01  &   ~690.6  $ \pm$ 1.3 & ~~4.16 $\pm $ 0.01\\
		$ K^{+}\pi^{+}\pi^{-}\pi^{-}\pi^{0}$  &  $(-0.057, 0.051)$ & 16.17  $\pm $   0.01  &   ~450.9  $ \pm$ 1.1 & ~~4.50 $\pm $ 0.02\\
\hline\hline
\end{tabular}
}
\label{tab:numST}
\end{center}
\end{table}

Candidates for the SL decay  are selected from the remaining tracks recoiling against the ST $\bar{D}^0$ mesons. 
Events containing a positron  candidate, with  a charge opposite to that of the charm quark in the ST $\bar{D}^0$ candidate, a $\pi^{-}$, and   a $\pi^0$  candidate, are accepted. Additionally, when there are multiple $\pi^0$ candidates, the one with the $\gamma\gamma$ invariant mass closest to the known $\pi^0$ mass~\cite{pdg16} is selected. To  suppress the backgrounds from the hadronic $D$ decays, the maximum energy of the unused showers ($E^{\rm max}_{\rm extra~ \gamma}$) is required to be less than 0.25~GeV, and no additional charged track ($N^{\rm char}_{\rm extra}$) or $\pi^0$  reconstructed from two unused photons ($N^{\pi^{0}}_{\rm extra}$) are allowed. 
 
The energy and momentum carried by the neutrino are denoted by $E_{\rm miss}$ and $\vec{p}_{\rm miss}$, respectively. They are calculated by $E_{\rm miss}=E_{\mathrm{beam}}-E_{\pi^-}-E_{\pi^0}-E_{e^+}$ and $\vec{p}_{\rm miss}=\vec{p}_{D^0}-\vec{p}_{\pi^-}-\vec{p}_{\pi^0}-\vec{p}_{e^+}$, in which $E_{\pi^-(\pi^0)(e^+)}$ and $\vec{p}_{\pi^-(\pi^0)(e^+)}$ are the energy and momentum of $\pi^-(\pi^0)(e^+)$ in the initial $e^+e^-$ rest frame. 
Furthermore, the momentum $\vec{p}_{D^0}$ is given by
$\vec{p}_{D^0} \equiv -\hat{p}_{\bar{D}^0}\sqrt{E_{\mathrm{beam}}^2/c^2-m^2_{\bar{D}^0}c^2},$
where $\hat{p}_{\bar{D}^0}$ is the unit vector of
momentum direction of the ST $\bar{D}^0$ and $m_{\bar{D}^0}$ is the known  $\bar{D}^0$ mass~\cite{pdg16}. 
Information on the undetected neutrino is obtained by using the variable $U_{\rm miss}$ defined by
\begin{equation}
U_{\rm miss} \equiv E_{\rm miss}-|\vec{p}_{\rm miss}|c ,
\end{equation}
which is expected to be zero.
The  backgrounds from the hadronic $D$ decays are  further suppressed with the requirement of $M_{\pi^{-}\pi^0e^+} <$ 1.70~GeV/$c^2$, where $M_{\pi^{-}\pi^0e^+}$ is the $\pi^{-}\pi^0e^+$ invariant mass.  To suppress the background from the decay of $D^0\to K^{-} e^+\nu_{e}$, we veto events with a $\pi^-\pi^0$ invariant mass within $\pm 70$ MeV/$c^2$ of the known kaon mass~\cite{pdg16}, which eliminates about 65.7\% of this background, while retaining about 98.6\% of the signal events.  This selection criterion is only imposed for the BF measurement, but not for the amplitude analysis. All requirements are obtained by optimizing the figure of merit  defined by $S/\sqrt{S+B}$, where $S$ and $B$ denote the signal and background yields from the  normalized inclusive MC sample.

\begin{figure}[tp!]
\begin{center}
   \flushleft
        \includegraphics[width=\linewidth]{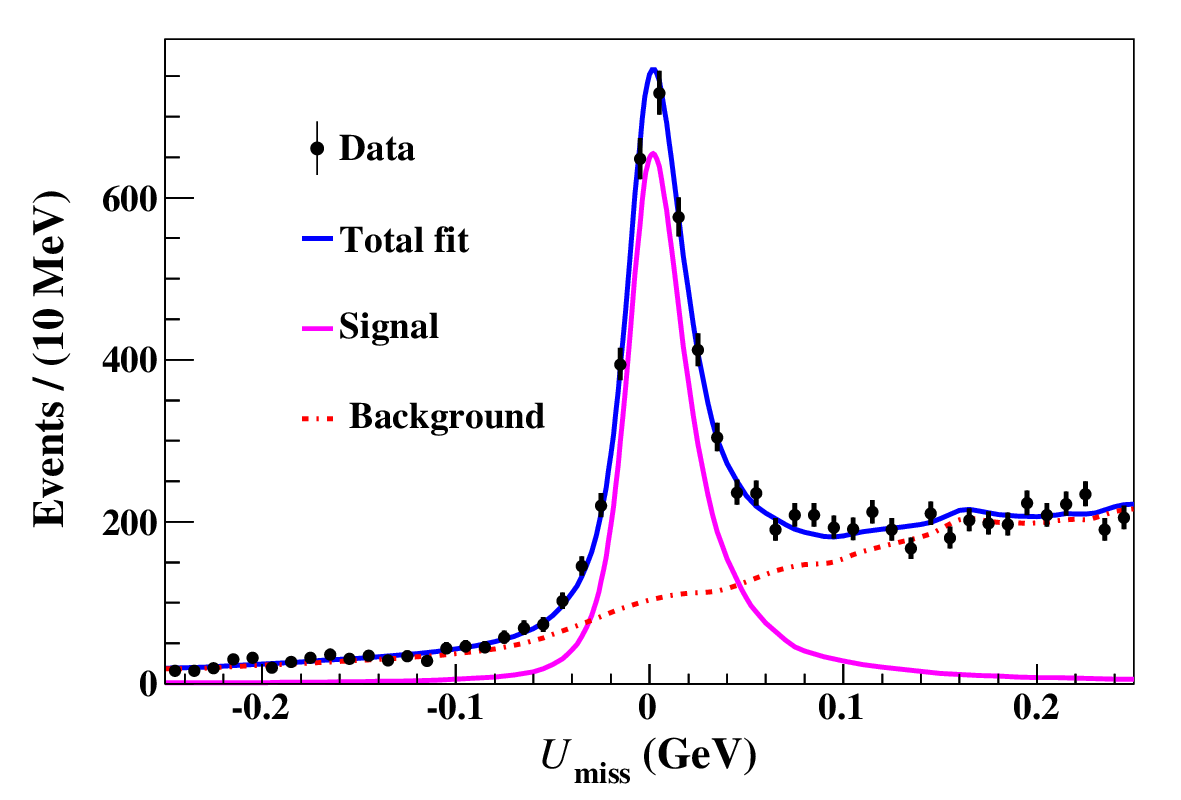}
   \caption{ Fit to the  $U_{\rm miss}$  distribution of the $D^{0}\to \pi^-\pi^{0}e^{+}\nu_{e}$ candidates. 
}
\label{fig:bf}
\end{center}
\end{figure}

 The $U_{\rm miss}$ distribution of the candidates passing the selection criteria is shown in Fig.~\ref{fig:bf}.
To obtain the signal yield, an unbinned maximum likelihood fit on this  distribution is performed.
In the fit, the signal is modeled by the MC-simulated shape extracted from the signal MC sample  convolved  with  a Gaussian function describing the difference of resolution between data and MC simulation.  The background is modeled by the MC-simulated shape obtained from the inclusive MC sample.   
The amplitude analysis (described later) shows  that the only significant resonance contribution to the $\pi^{-}\pi^0$ system is $D^0\to \rho(770)^-e^+\nu_e$. 
 The DT efficiencies with different tag modes  are summarized in Table~\ref{tab:numST}.
 Using  Eq.~(\ref{eq:branch}),  with a signal yield of $N_{\rm DT}=3337\pm77({\rm stat.})$, the detection efficiency $\varepsilon_{\rm SL}$ is estimated to be $(29.72\pm0.03)\%$ ($\varepsilon_{\rm SL}$ is corrected according to the discussions in the systematic uncertainty part of the BF), and the corresponding BF is determined as $\mathcal B({D^0\to \rho(770)^-e^+\nu_e})=(1.439\pm0.033({\rm stat.})\pm 0.027(\rm syst.))\times 10^{-3}$. 
 
When measuring the BF using the DT approach, many systematic uncertainties due to the ST selection cancel. The remaining sources of systematic uncertainty are discussed below.
\begin{itemize}
\item $N^{\rm tot}_{\rm ST}$. The systematic uncertainty of the fits to the $M_{\rm BC}$ spectra is assigned to be 0.1\%~\cite{prd109_2024}, after examining the relative change in the ST yield between data and MC simulation after  varying the signal shape and the endpoint of the ARGUS function within $\pm 0.2$ MeV for the background shape.

\item  $\pi^-$ tracking and PID  efficiencies. The uncertainties from the tracking and PID  efficiencies of  $\pi^-$  are assessed by  analyzing DT control sample of   $\psi(3770)\to D\bar{D}$ events with hadronic $D$ decays~\cite{prl127_131801, prd102_112005}. The momentum-weighted data-MC differences are 0.997 $\pm$ 0.005  and  0.998 $\pm$ 0.005  arising from the  $\pi^-$ tracking and PID efficiencies, respectively. The signal efficiencies applied to data are corrected by these factors.
 After these corrections, we assign 0.5\% and 0.5\% as the systematic uncertainty for $\pi^-$ tracking and $\pi^-$ PID, respectively.

\item $e^+$  tracking and PID  efficiencies. The uncertainties from the positron tracking and PID  efficiencies  are studied with a control sample of $e^+e^-\to \gamma e^+e^-$ events~\cite{prl127_131801, prd102_112005}. The data-MC differences are 1.002 $\pm$ 0.005 for the $e^+$ tracking efficiency and 0.972  $\pm$  0.005 for the  $e^+$ PID efficiency. We correct the signal efficiency in data  by these factors and assign 0.5\% and 0.5\% as the systematic uncertainty for $e^+$ tracking and $e^+$ PID, respectively.
 
\item $\pi^0$ reconstruction.  The uncertainty of  the $\pi^0$ reconstruction is studied with a DT control sample of  $D^0\to K^-\pi^+\pi^0$ decays. After correcting  the differences of the $\pi^0$ reconstruction efficiencies  between data and MC simulation, which is $0.993 \pm 0.008$, we take 0.8\% as the systematic uncertainty.

\item $K$ rejection. The efficiency of the $K$ rejection requirement is greater than 98\% and the difference in these efficiencies between data and MC simulation is negligible. 

\item  $E^{\rm max}_{\rm extra \gamma}$,  $N_{\rm extra}^{\rm char}$ and $N_{\rm extra}^{\pi^0}$. The uncertainty associated with the $E^{\rm max}_{\rm extra \gamma}$,  $N_{\rm extra}^{\rm char}$ and $N_{\rm extra}^{\pi^0}$  requirements is estimated  by analyzing a DT sample of $D^0\to K^-\pi^{0} e^+\nu_e$ decays.
The difference in efficiency of these requirements between data and MC simulation is $1.018 \pm 0.005$, which we apply as a correction with associated systematic  uncertainty.  

\item $M_{\rm \pi^-\pi^0e^{+}}$. The uncertainty associated with the $M_{\rm \pi^-\pi^0e^{+}}$   requirement is estimated  by analyzing the DT sample of  $D^0\to K^-\pi^{0} e^+\nu_e$ decays.
The difference in the efficiency between data and MC simulation is $1.007 \pm 0.003$, which we apply as a correction with associated systematic uncertainty.  

\item  ${ U_{\rm miss}}$ fit. The uncertainty associated with the fit of the $U_{\rm miss}$ distribution is estimated by varying the relative fraction of the major background from $e^+e^- \to D^{0}\bar{D}^{0}$ events within the uncertainty of its cross section, and the dominant background channels within the uncertainty of  their input BFs in the inclusive MC sample. The change seen with respect to the baseline BF result, 0.2\%, is assigned as the corresponding systematic uncertainty.

\item MC model.  The uncertainty related to the signal MC model is estimated  by comparing the signal efficiencies with variations in the input form-factor parameters by $\pm 1\sigma$.   The largest change in the signal efficiency, 0.6\%, from this variation is taken as the corresponding uncertainty.

\item MC sample size. The uncertainty due to the limited MC sample size, 0.2\%, is assigned as a systematic uncertainty. 

\item  Input BF. The uncertainty due to the assumed BF of  $\pi^0 \to \gamma \gamma$ is  0.1\%~\cite{pdg16}.
\end{itemize}

The above sources are summarized in Table~\ref{tab:bf-syst-sum}. The systematic uncertainty contributions are summed in quadrature, and the total systematic uncertainty on the BF measurement is 1.9\%.

\begin{table}[htp]
\centering
\caption{Relative systematic uncertainties  in the BF measurement. 
}
\begin{tabular}{lcc|c}
  \hline
  \hline
  Source                              &Uncertainty (\%)\\
  \hline
  $N^{\rm tot}_{\rm ST}$              & 0.1       \\
  $\pi^-$ tracking                   & 0.5       \\
  $\pi^-$ PID                  & 0.5       \\
  $e^+$  tracking                         & 0.5       \\
  $e^+$  PID                         & 0.5      \\
  $\pi^0$ reconstruction                   & 0.8       \\
 $K$ rejection                          & Neglected      \\
  $E^{\rm max}_{\rm extra \gamma}$,  $N_{\rm extra}^{\rm char}$ and $N_{\rm extra}^{\pi^0}$       & 0.5 \\
 $M_{\rm \pi^-\pi^0e^{+}}$                            & 0.3       \\
  ${ U_{\rm miss}}$ fit              & 0.2       \\
  MC model                       & 0.6       \\
  MC sample size                           & 0.2       \\
  Input BF                           & 0.1       \\
  \hline
  Total                               & 1.9       \\

  \hline
  \hline
\end{tabular}
\label{tab:bf-syst-sum}
\end{table}

\section{Amplitude Analysis OF $D^0 \to \pi^- \pi^0 e^+ \nu_e$ Decay}
To increase the signal purity for the amplitude analysis, we require $|{ U_{\rm miss}}|<0.03$ GeV. Additionally, a requirement of $p_{\pi^0} >0.1$ GeV/$c$ is used to suppress the  background from $D^0\to K^{-} e^+\nu_{e}$, in which  the $K^{-}$ decays to $\pi^- \pi^0$.
An additional source of background comes from events containing $D^+ \to K^0_{S}(\pi^0 \pi^0)  e^+ \nu_e$ versus $D^- \to K^{+}\pi^{-}\pi^{-}$ or $D^- \to K^{+}\pi^{-}\pi^{-}\pi^{0}$, or events containing $D^+ \to  K^0_{S}(\pi^- \pi^+)  e^+ \nu_e$ versus $D^- \to K^{+}\pi^{-}\pi^{-}\pi^{0}$ or $D^- \to K^0_{S}\pi^{-}\pi^{0}$.
Such events can potentially fake  $D^0 \to \pi^- \pi^0 e^+ \nu_e$ versus  $\bar{D}^0 \to K^{+}\pi^{-}\pi^{0}$, $\bar{D}^0 \to K^{+}\pi^{-}\pi^{0}\pi^{0}$, $\bar{D}^0 \to K^{+}\pi^{+}\pi^{-}\pi^{-}$, or $\bar{D}^0 \to K^0_{S}\pi^{+}\pi^{-}$ due to the exchange of the $\pi^0$($\pi^+$) from the $D^+$ decay and the $\pi^-$($\pi^0$) from the $D^-$ decay.  We reconstruct events under these background hypotheses and calculate the  $D^-$ invariant mass.  Events are rejected if the mass of the $D^-$ candidate lies within $\pm 30$ MeV/$c^2$ of the known $D^-$ mass.    This selection leads to a sample containing 2075  signal events with a background fraction of  $f_b$ = (18.7~$\pm$~0.8)\%.

 The differential decay width of $D^0 \to \pi^- \pi^0 e^+ \nu_e$ can be expressed in terms of five kinematic variables~\cite{prd46_5040}:
the square of the invariant mass of the $\pi^-\pi^0$ system ($m^2$), the square of the invariant mass of the $e^+ \nu_e$ system ($q^2$), the angle between the momentum of the $\pi^-$  in the $\pi^-\pi^0$ rest frame and the momentum of the $\pi^-\pi^0$ system in the $D^0$ rest frame ($\theta_{\pi^-}$), the angle between the momentum of the $e^+$ in the $e^+\nu_e$ rest frame and the momentum of the $e^+\nu_e$ system in the $D^0$ rest frame ($\theta_{e^{+}}$), and the angle between the normal of the decay plane of the $e^+\nu_e$  pair and that of the $\pi^-\pi^0$ pair ($\chi$), both defined in the $D^0$ rest frame .
Neglecting the mass of the positron, the differential decay width of $D^0 \to \pi^- \pi^0 e^+ \nu_e$ can be expressed as~\cite{prd46_5040}
\begin{eqnarray}
{\rm d}^5\Gamma&=&\frac{G^{2}_{F}|V_{cd}|^{2}}{(4\pi)^{6}m^{3}_{D^0}}X\beta {\cal I}(m^{2}, q^{2}, \theta_{\pi^-}, \theta_{e^{+}}, \chi) \nonumber \\
         && {d}m^{2} {d}q^{2}{d}\cos\theta_{\pi^-}{d}\!\cos\theta_{e^{+}}{d}\chi.
         \label{eq:pwa}
\end{eqnarray}
Here, $X=p_{\pi^- \pi^0}m_{D^0}$, where $p_{\pi^- \pi^0}$ is the momentum of $\pi^- \pi^0$ in the $D^0$ rest frame, and $\beta=2p^{*}/m$, with $p^{*}$ denoting the momentum of $\pi^-$ in the $\pi^- \pi^0$ rest frame.
 The Fermi coupling constant is denoted by $G_F$ and $V_{cd}$ is the $c\to d$ element of the Cabibbo-Kobayashi-Maskawa matrix.
The dependence of the decay density $\mathcal{I}$ is given by
\begin{eqnarray}
\mathcal{I}&=&\mathcal{I}_1+\mathcal{I}_2{\rm cos2}\theta_e+\mathcal{I}_3{\rm sin}^2\theta_e{\rm cos}2\chi+\mathcal{I}_4{\rm sin}2\theta_e{\rm cos}\chi  \nonumber\\
           &+&\mathcal{I}_5{\rm sin}\theta_e{\rm cos}\chi+\mathcal{I}_6{\rm cos}\theta_e+\mathcal{I}_7{\rm sin}\theta_e{\rm sin}\chi \nonumber \\
           &+&\mathcal{I}_8{\rm sin}2\theta_e{\rm sin}\chi+\mathcal{I}_9{\rm sin}^2\theta_e{\rm sin}2\chi,
\label{eq:Ifunc}
\end{eqnarray}
where $\mathcal{I}_{1, \ldots,9}$ depend on  $m^2$, $q^2$ and $\theta_{\pi^-}$. These 
quantities can be expressed in terms of the three form factors $\mathcal{F}_{1,2,3}$:
\begin{equation}
  \begin{aligned}
  \mathcal{I}_{1}=&\frac{1}{4}\{|\mathcal{F}_{1}|^{2}+\frac{3}{2}\sin^{2}\theta_{\pi^-}
  (|\mathcal{F}_{2}|^{2}+|\mathcal{F}_{3}|^{2})\}, \\
  \mathcal{I}_{2}=&-\frac{1}{4}\{|\mathcal{F}_{1}|^{2}-\frac{1}{2}\sin^{2}\theta_{\pi^-}
  (|\mathcal{F}_{2}|^{2}+|\mathcal{F}_{3}|^{2})\}, \\
  \mathcal{I}_{3}=&-\frac{1}{4}\{|\mathcal{F}_{2}|^{2}-|\mathcal{F}_{3}|^{2}\}\sin^{2}\theta_{\pi^-}, \\
  \mathcal{I}_{4}=&\frac{1}{2}\rm{Re}(\mathcal{F}^{*}_{1}\mathcal{F}_{2})\sin\theta_{\pi^-}, \\
  \mathcal{I}_{5}=&\rm{Re}(\mathcal{F}^{*}_{1}\mathcal{F}_{3})\sin\theta_{\pi^-}, \\
  \mathcal{I}_{6}=&\rm{Re}(\mathcal{F}^{*}_{2}\mathcal{F}_{3})\sin^{2}\theta_{\pi^-}, \\
  \mathcal{I}_{7}=&\rm{Im}(\mathcal{F}_{1}\mathcal{F}_{2}^{*})\sin\theta_{\pi^-}, \\
  \mathcal{I}_{8}=&\frac{1}{2}\rm{Im}(\mathcal{F}_{1}\mathcal{F}_{3}^{*})\sin\theta_{\pi^-}, \\
  \mathcal{I}_{9}=&-\frac{1}{2}\rm{Im}(\mathcal{F}_{2}\mathcal{F}^{*}_{3})\sin^{2}\theta_{\pi^-}.
  \end{aligned}
\end{equation}
Then one can expand $\mathcal{F}_{i=1,2,3}$ into partial waves including
 $S$-wave ($\mathcal{F}_{10}$) and $P$-wave ($\mathcal{F}_{i1}$).
 Consequently, the form factors can be written as
\begin{equation}
  \begin{aligned}
  {\cal F}_{1}&={\cal F}_{10}+{\cal F}_{11}\cos\theta_{\pi^-},~  
  {\cal F}_{2}&=\frac{1}{\sqrt{2}}{\cal F}_{21}, ~
  {\cal F}_{3}&=\frac{1}{\sqrt{2}}{\cal F}_{31}. ~
  \label{eq:form_factor}
  \end{aligned}
\end{equation}
The $P$-wave related form factors $\mathcal{F}_{i1}$ are parameterized by 
the helicity basis form factors $H_{0,\pm}$: 
\begin{equation}
  \begin{aligned}
  \mathcal{F}_{11}=&2\sqrt{2}\alpha qH_{0}\times \mathcal{A}(m),  \\
  \mathcal{F}_{21}=&2\alpha q(H_{+}+H_{-})\times \mathcal{A}(m), \\
  \mathcal{F}_{31}=&2\alpha q(H_{+}-H_{-})\times \mathcal{A}(m). \\
  \label{eq:form_factor_P}
  \end{aligned}
\end{equation} 
Here $\mathcal{A}(m)$ denotes the amplitude characterizing the shape
of the resonances, which follows a Gounaris-Sakurai form~\cite{prl21_244}  defined in Eq.~(\ref{eq:Am}).
 The constant factor $\alpha$, defined in Ref.~\cite{prd94_032001}, depends on the definition of $\mathcal{A}(m)$.
The helicity basis form factors can be related to 
one vector $V(q^{2})$ and two axial-vector $A_{1,2}(q^{2})$
form factors:
\begin{equation}
  \begin{aligned}
  H_{0}(q^{2},m^{2})=&\frac{1}{2mq}[(m_{D^0}^{2}-m^{2}-q^{2})(m_{D^0}+m)A_{1}(q^{2}) \\
  &-4\frac{m_{D^0}^{2}p_{\pi^- \pi^0}^{2}}{m_{D^0}+m}A_{2}(q^{2})], \\
  H_{\pm}(q^{2},m^{2})&=[(m_{D^0}+m)A_{1}(q^{2})\mp\frac{2m_{D^0}p_{\pi^- \pi^0}}{(m_{D^0}+m)}V(q^{2})].
  \label{eq:helicity}
  \end{aligned}
\end{equation}
We use the  single-pole model~\cite{Dstokkmunu} to describe the $q^{2}$ dependence:
\begin{equation}
  \begin{aligned}
      V(q^{2})&=\frac{V(0)}{1-q^{2}/m_{V}^{2}},  \\
  A_{1}(q^{2})&=\frac{A_{1}(0)}{1-q^{2}/m_{A}^{2}},  \\
  A_{2}(q^{2})&=\frac{A_{2}(0)}{1-q^{2}/m_{A}^{2}},  
  \label{eq:spd}
  \end{aligned}
\end{equation}
where $m_{V}$ and $m_{A}$ are the pole masses and fixed to $m_{D^{*}(2010)}=$ 2.01  GeV/$c^{2}$  and $m_{D_{1}(2420)}=$ 2.42 GeV/$c^{2}$~\cite{pdg16}  in the fit, respectively.
 Therefore, it is natural to define the two coupling constants, $r_{V}=V(0)/A_{1}(0)$ and $r_{2}=A_{2}(0)/A_{1}(0)$ as the hadronic form-factor ratios at the momentum square $q^{2}=0$.  They are determined from the amplitude analysis fit.

The amplitude of the $P$-wave resonance $\mathcal{A}(m)$ is expressed as a  Gounaris-Sakurai  function~\cite{prl21_244} to describe the $\rho(770)^-$ line shape,
\begin{equation}
\mathcal{A}(m)= \frac{m^2_{0}(1+\Gamma_{0}g/m_{0})(p^{*}/{p^{*}_{0})}}{m_{0}^{2}-m^{2}+ f(m)-im_{0}\Gamma(m)}\frac{B(p^{*})}{B(p^{*}_{0})}, 
\label{eq:Am}
\end{equation}
where $B(p) = 1/\sqrt{1+r_{\rm BW}^{2}p^{2}}$ with $r_{\rm BW}$ = 3.07 $({\rm GeV}/c)^{-1}$ and $\Gamma(m)=\Gamma_{0} \left(\frac{p^{*}}{p^{*}_{0}}\right)^{3}\left(\frac{m_{0}}{m}\right)$, where   $p^{*}_{0}$ is the modulus of the momentum of the $\pi^-$  at the pole mass of the resonance $m_0$. The parameter $g$ and  the function $f(m)$ are taken from Ref.~\cite{prl21_244}; $m_{0}$ and $\Gamma_{0}$ are the pole  mass and total width of the resonance, respectively. 

The $S$-wave related $\mathcal{F}_{10}$ is expressed as
\begin{equation}
 \mathcal{F}_{10}=p_{\pi^-\pi^{0}}m_{D^0}\frac{1}{1-\frac{q^{2}}{m_{A}^{2}}}\mathcal{A}_{S}(p^{*}), \\
\end{equation}
where the expression of the amplitude  $\mathcal{A}_{S}(p^{*})=\frac{1}{1-i a_{0} p^{*}}$ is adopted, with  $a_{0}= (-0.11\pm 0.01)m^{-1}_{\pi}$~\cite{Swavelass}.

\begin{figure*}[tp!]
\begin{center}
      \includegraphics[width=0.31\linewidth]{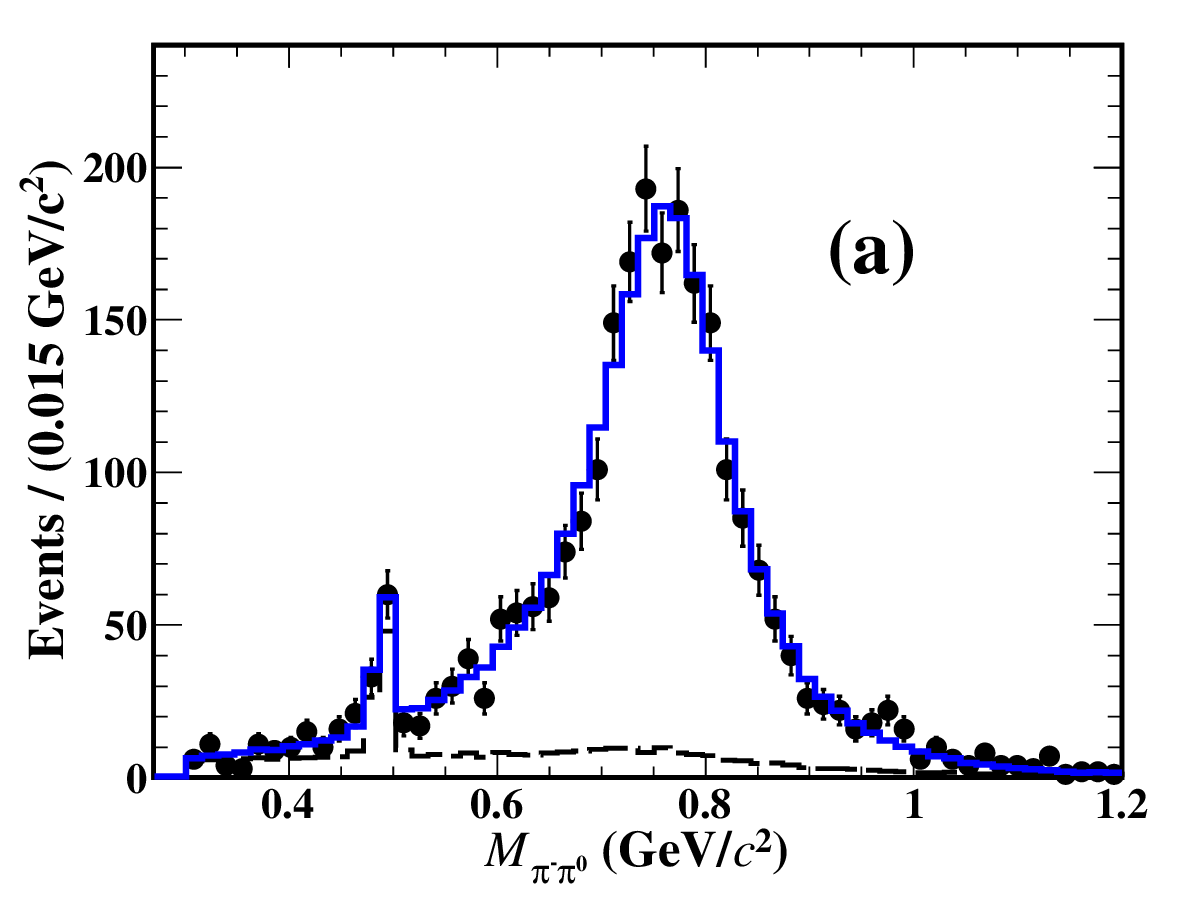}
      \includegraphics[width=0.31\linewidth]{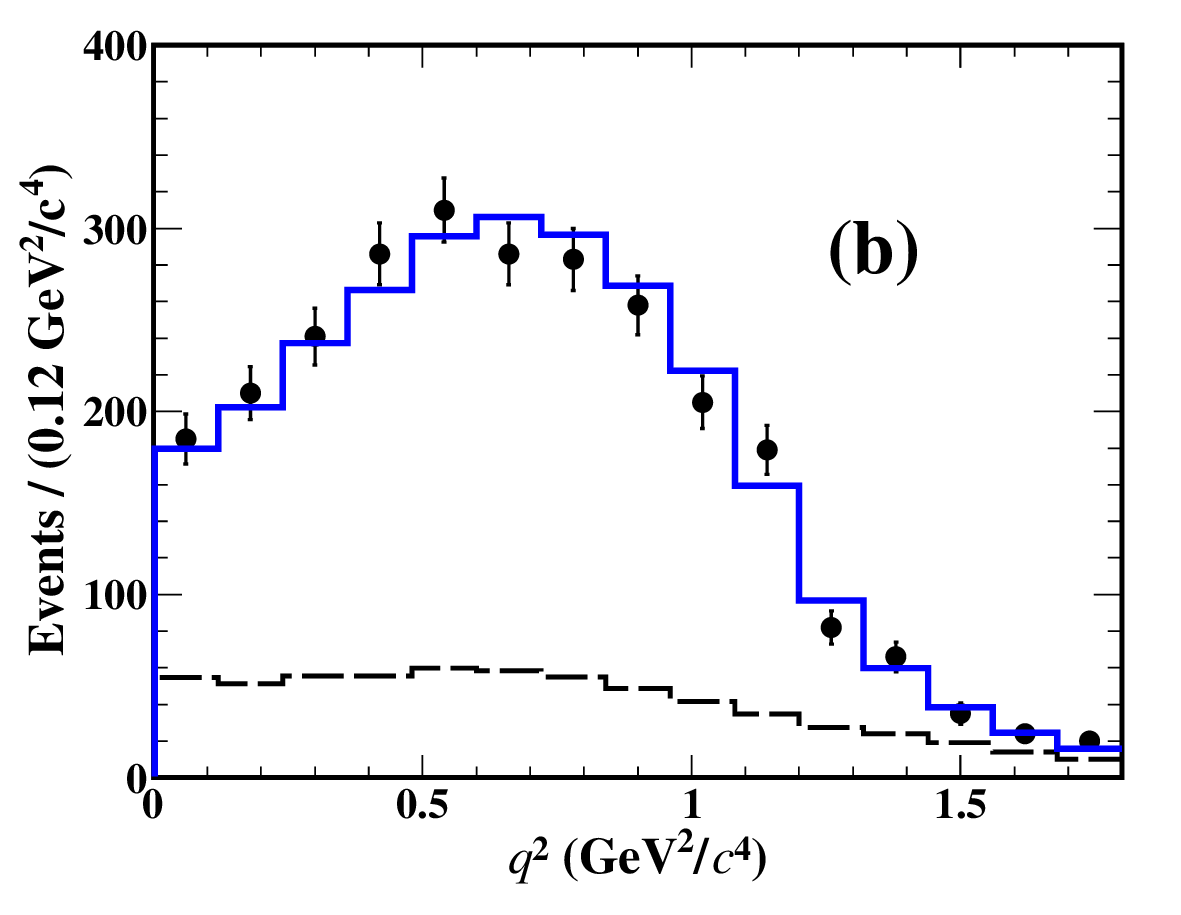}
       \includegraphics[width=0.31\linewidth]{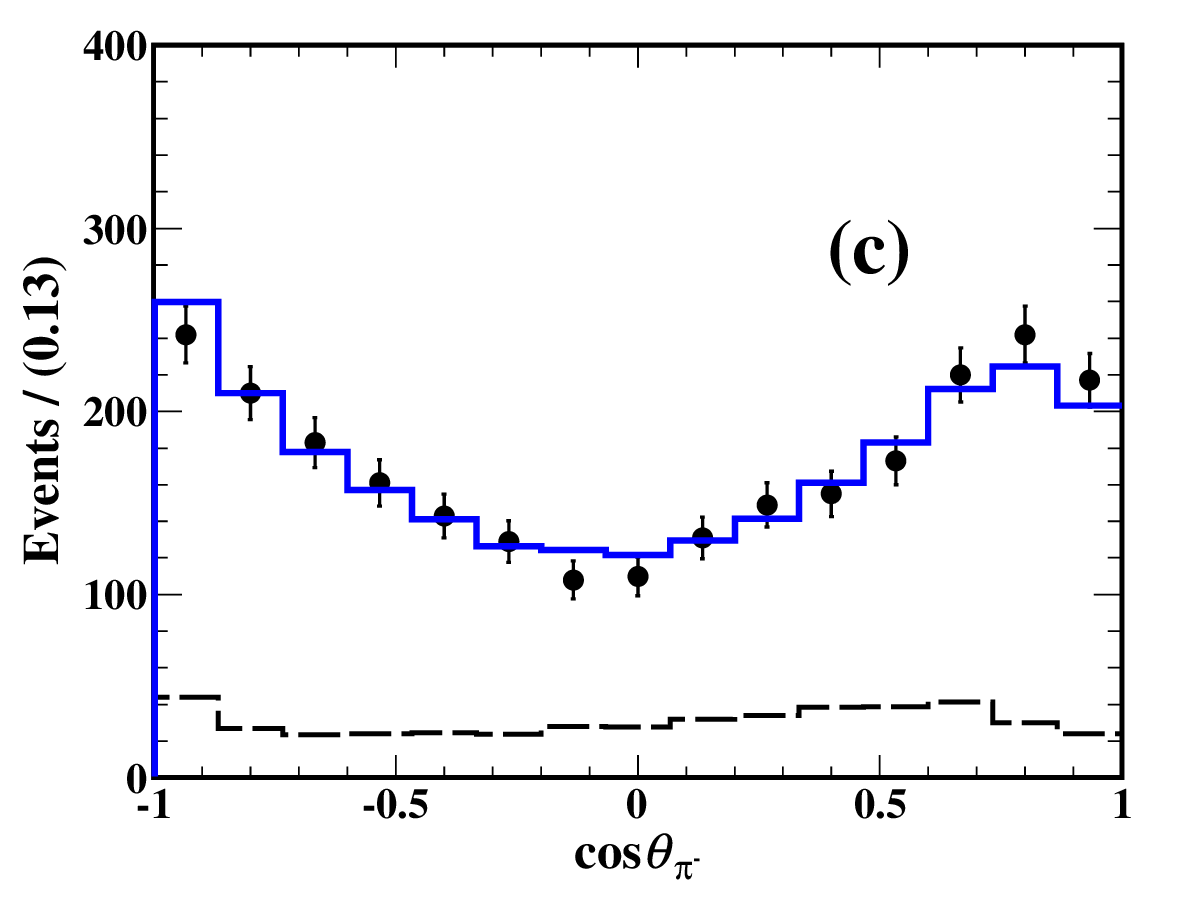}
      \includegraphics[width=0.31\linewidth]{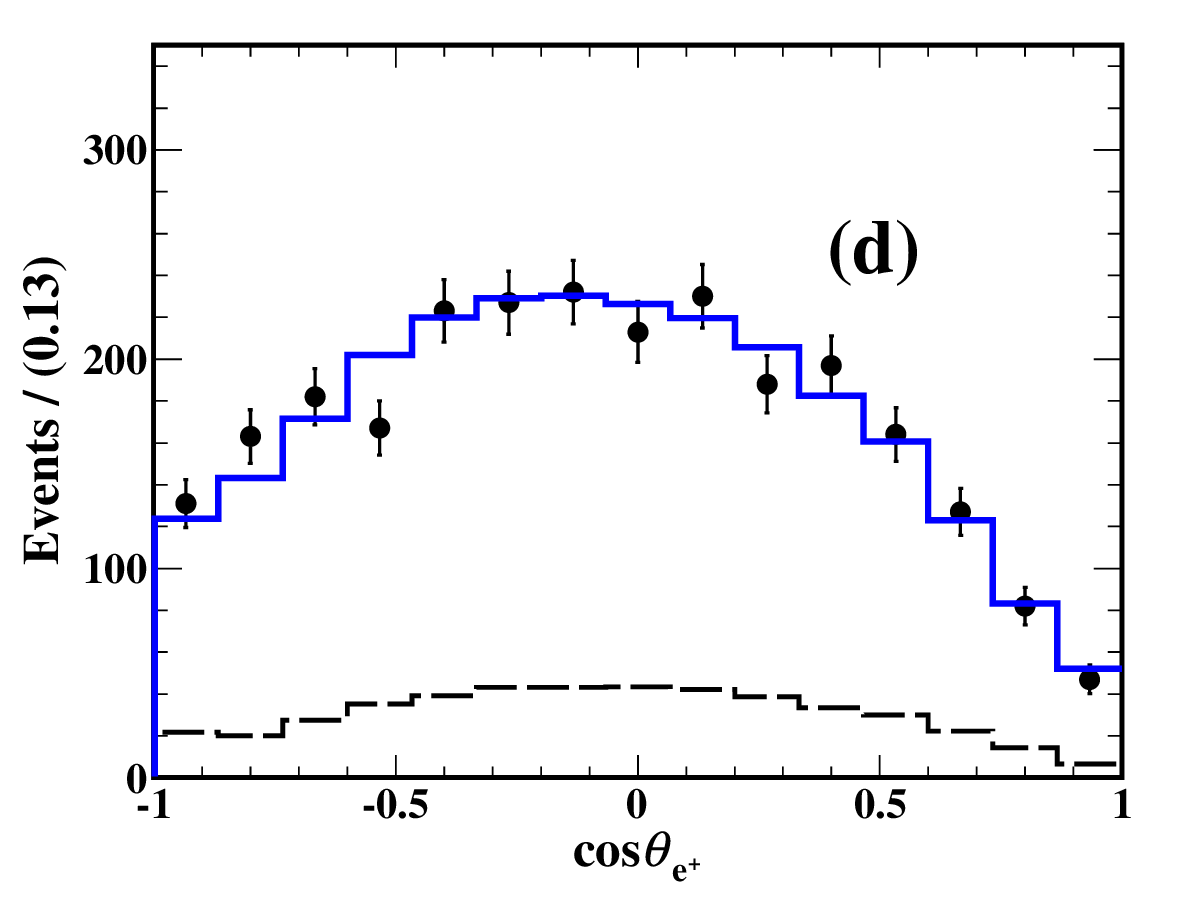}
      \includegraphics[width=0.31\linewidth]{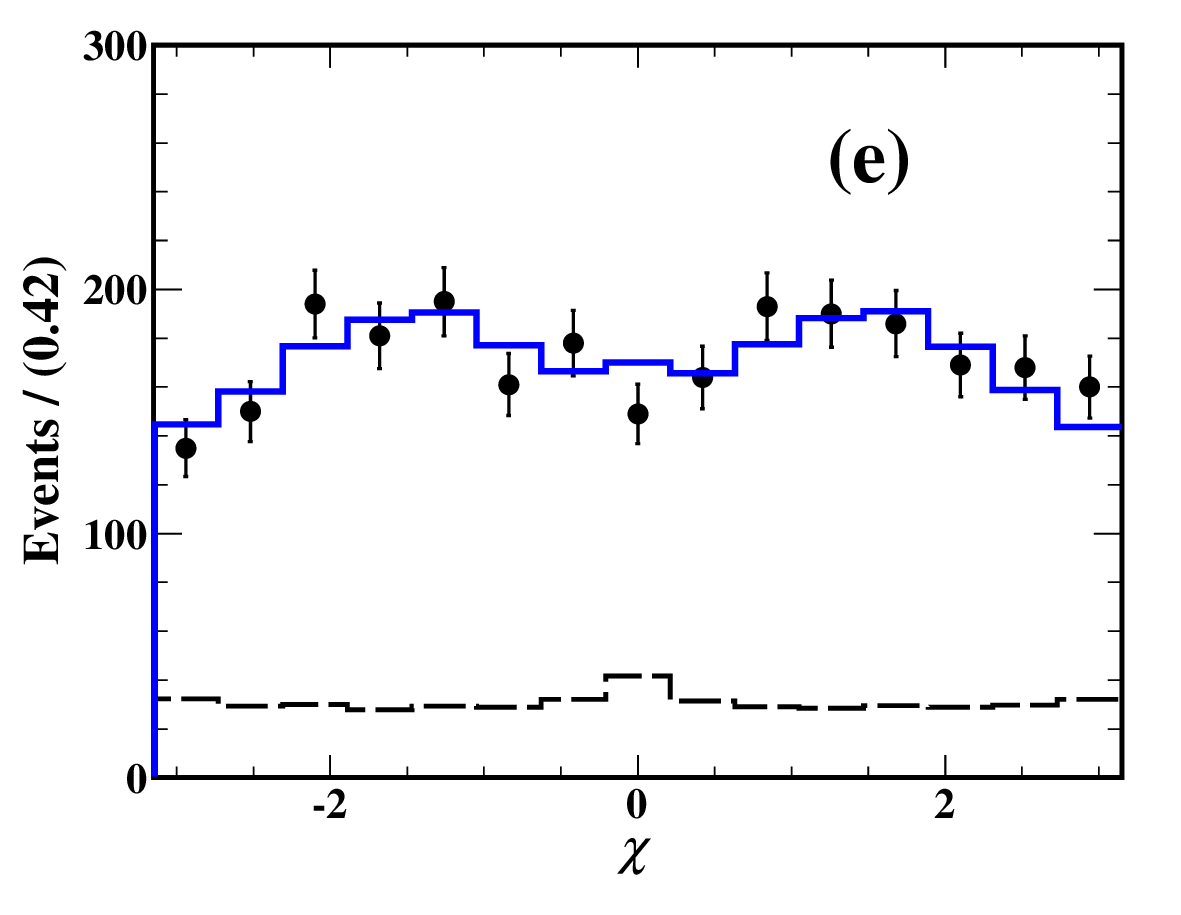}
   \caption{Projections of the amplitude analysis on   (a) $M_{\pi^-\pi^0}$, (b) $q^2$,   (c) $\cos\theta_{\pi^-}$, (d) $\cos\theta_{e^+}$, and  (e)  $\chi$. 
     The dots with error bars are data, the blue lines are  signal MC with a decay model determined by the amplitude analysis, and the dashed lines are the simulated background.  The peak around 0.5 GeV/$c^2$ in the ${ M}_{\pi^{-}\pi^{0}}$ distribution comes from the $D^0\to K^{-}e^{+}\nu_e$ background.
   }
\label{fig:formfactor}
\end{center}
\end{figure*}

The  amplitude analysis is performed using an unbinned maximum likelihood fit.
The negative log likelihood $-\ln\!{\cal L} $ is defined as~\cite{prd85_122002}
\begin{equation} 
\begin{aligned}
 -\ln\!{\cal L} &=  -\sum_{i=1}^{N}\ln  \left[(1-f_{b}) \frac{\omega(\xi_{i},\eta)}{\int \omega(\xi_{i},\eta) \, \epsilon(\xi_{i}) \, R_4(\xi_{i}) \, {\rm d}\xi_{i}} \right. \\
 &\quad \left. + f_{b}\frac{B_{\epsilon}(\xi_{i})}{\int B_{\epsilon}(\xi_{i}) \, \epsilon(\xi_{i}) \, R_4(\xi_{i}) \,  {\rm d}\xi_{i} }\right],
  \label{eq:lnL}
\end{aligned}
\end{equation}
where $\xi_{i}$ denotes the five kinematic variables characterizing of one event  and $\eta$ denotes the fit parameters  $r_V$ and $r_2$; $\omega(\xi_{i},\eta)$ is the decay intensity [i.e., $\mathcal{I}$ in Eq.~(\ref{eq:pwa})]; $B_{\epsilon}(\xi_{i})=B(\xi_{i})/\epsilon(\xi_{i})$ is the efficiency-corrected background shape,  where $B(\xi_{i})$ is a function that describes the background, $\epsilon(\xi_{i})$ is the reconstruction efficiency for the final state $\xi_{i}$, and $R_4(\xi_{i})$ is an element of four-body phase space.
 
 The normalization integral terms can be written as
\begin{equation}
  \begin{aligned}
\int \omega(\xi_{i},\eta) \, \epsilon(\xi_{i}) \, R_4(\xi_{i}) \, {\rm d}\xi_{i}
      \propto\frac{1}{N_{\rm selected}}\sum_{k=1}^{N_{\rm selected}} \frac{\omega(\xi_{k}, \eta)}{\omega(\xi_{k}, \eta_{0})}, \\
\int B_{\epsilon}(\xi_{i}) \, \epsilon(\xi_{i}) \, R_4(\xi_{i}) \,  {\rm d}\xi_{i} 
      \propto\frac{1}{N_{\rm selected}}\sum_{k=1}^{N_{\rm selected}} \frac{ B_{\epsilon}(\xi_{k})}{\omega(\xi_{k}, \eta_{0})}.
\label{eq:sigmc-integral}
  \end{aligned}
\end{equation}
Here the terms $\eta$ and $\eta_{0}$ represent the values of the parameters used in the fit 
and those used to produce the simulated events, respectively.
$N_{\rm selected}$ denotes the number of the signal MC events after reconstruction and selection. 

 The background shape is derived from the inclusive MC sample, and its fraction $f_b$ is fixed according to the result of the $U_{\rm miss}$ fit.  The value of $\epsilon(\xi_i)$ is obtained by calculating the ratio between the numbers of selected and truth events using phase-space MC samples, which are divided into 8 × 8 × 8 × 8 × 8 bins in the five-variable space $(m^{2}, q^{2}, \theta_{\pi^-}, \theta_{e^{+}}, \chi)$. For some edge bins, we merge neighboring bins until a minimum of twenty events are accumulated~\cite{Dstokkmunu}.  

 The structure of the $\pi^-\pi^0$ system is dominated by the vector meson $\rho(770)^-$; nevertheless, the contribution from the $S$-wave has been considered but the fitting result indicates that the statistical significance of this component is zero.
 Therefore, only the $\rho(770)^-$ in the $\pi^-\pi^0$ system is considered in the baseline solution. 
In the fit,  the mass and width of  $\rho(770)^-$  are fixed to the PDG values~\cite{pdg16}. The hadronic form-factor ratios $r_{V}=\frac{V(0)}{A_{1}(0)}=1.548\pm0.079(\rm stat.)\pm0.041(\rm syst.)$ and $r_{2}=\frac{A_{2}(0)}{A_{1}(0)}=0.823\pm0.056(\rm stat.)\pm0.026(\rm syst.)$ are obtained, with a correlation coefficient $\rho_{r_V,r_2}=-0.21$.
The projected distributions of the fit onto the fitted variables are shown in Fig.~\ref{fig:formfactor}. 
The fitting process is validated using a large simulated sample of inclusive events, where the pull distributions of $r_{V}$ and $r_{2}$ are found to be consistent with a normal distribution.

\begin{table}[htp]
\begin{center}
\caption{ Relative  systematic uncertainties (in \%) of the hadronic form-factor ratio measurements. }
\vspace{0.25cm}
\begin{tabular}{p{4.0cm}lccc}
\hline
\hline
Source                        & $\triangle r_V$  &  $\triangle r_2$\\
\hline
Background   fraction                &  0.22  &  0.71  \\
Background   shape                  &  0.66  &  0.85   \\
$r_{\rm BW}$                                                  &  0.35  &  0.93\\
$m_V$                                                      &  1.88  &  0.03 \\
$m_A$                                                      &  1.65  &  2.85 \\
$\rho(770)^-$ line shape                           &  0.01  &  0.01\\
$S$-wave  component                              &  0.01  &  0.05\\
Efficiency corrections                                 &  0.07 &  0.12 \\
\hline 
Total                                                           & 2.63 & 3.20  \\
\hline
\hline
\end{tabular}
\label{tab:ff-syst-sum}
\end{center}
\end{table}

The following sources of systematic uncertainties, summarized in Table~\ref{tab:ff-syst-sum}, have been considered in the measurement of the hadronic form-factor ratios.
\begin{itemize}
\item Background  estimation.
The systematic uncertainties due to the background  fraction and background shape  are estimated by varying the background fraction  $f_{b}$  and  varying the cross section of the dominant background from $e^+e^- \to D^{0}\bar{D}^{0}$ by $\pm 1\sigma$~\cite{corss_section}, respectively.  The differences caused by these variations are assigned as the uncertainties. 

\item $r_{\rm BW}$. The systematic uncertainty in the fixed parameter of  $r_{\rm BW}$ is estimated by varying their input values by $\pm 1\sigma$~\cite{prd94_032001}, taking the largest difference with respect to the baseline  result as the systematic uncertainties.

\item $m_V$ and $m_A$. The systematic uncertainties in the fixed parameters of $m_V$ and $m_A$ are estimated by varying their input values by $\pm 100$~MeV/$c^2$~\cite{Dstokkmunu}. The differences with respect to the baseline result are assigned as the systematic uncertainties.  

\item $\rho(770)^-$ line shape. The uncertainty in the $\rho(770)^-$ line shape is estimated by varying the mass and width of $\rho(770)^-$  by $\pm 1\sigma$~\cite{pdg16}. The largest difference is taken as the systematic uncertainty.

\item $S$-wave  component. The systematic uncertainty due to neglecting a possible contribution from the $S$-wave component is estimated by incorporating the $S$-wave component in Eq.~(\ref{eq:form_factor}). The difference with respect to the baseline result is assigned as the systematic uncertainty.

\item Efficiency corrections.  To estimate the systematic uncertainty  related to the reconstruction efficiency, the fit is performed by varying the PID and  tracking efficiencies according to their uncertainties.  The difference with respect to the baseline result is assigned as the systematic uncertainty. 
\end{itemize}

The total systematic uncertainty for each case is calculated by summing all the individual contributions listed in Table~\ref{tab:ff-syst-sum} in quadrature.

\section{Summary}
In summary, we present  the study of the dynamics  for the SL decay of $D^0\to \pi^-\pi^0e^{+}\nu_{e}$ by analyzing an $e^+e^-$ annihilation data sample of 7.93 fb$^{-1}$ collected at the center-of-mass energy of 3.773 GeV with the BESIII detector. The  BF  of $D^0\to \rho(770)^-e^+\nu_e$ is measured as $(1.439 \pm 0.033(\rm stat.) \pm  0.027(\rm syst.))\times 10^{-3}$. It is consistent with the previous BESIII measurement~\cite{pipienubesiii}, but with a 1.6 factor  improvement in precision. In addition, the hadronic form-factor ratios of the $D^0\to \rho^-(770)^-e^+\nu_e$ decay are determined to be $r_V = 1.548 \pm 0.079(\rm stat.) \pm 0.041(\rm syst.)$ and $r_2 = 0.823 \pm 0.056(\rm stat.) \pm 0.026(\rm syst.)$. 
A comparison of our results with previous measurements is summarized in Table~\ref{tab:com}. The findings are consistent within uncertainties with the previous BESIII and CLEO measurements of $D\to \rho e^+\nu_e$~\cite{pipienubesiii,pipienucleo}. 
The measured form-factor ratios $r_V$ and $r_2$, support the predictions of the CQM~\cite{cqm_2000}, the CCQM~\cite{fpb14_66401}, and the LFQM~\cite{lfqm_2012} calculations. Conversely, they disfavor the LCSR~\cite{Ijmp21_6125-6172} and the HM$\chi$T model~\cite{hmt_2005}.

\begin{table}[htp]
\begin{center}
\caption{  Comparison of $r_V$ and $r_2$ measured in this paper with previous measurements of $D\to \rho e^+\nu_e$.
}
\vspace{0.25cm}
\begin{tabular}{lccc}
\hline
Experiments                        & $r_V$                       & $r_2$                       \\
\hline
This analysis                      &  1.548$\pm$0.079$\pm$0.041    & 0.823$\pm$0.056$\pm$0.026   \\
\hline
BESIII~\cite{pipienubesiii}        &  1.695$\pm$0.083$\pm$0.051  & 0.845$\pm$0.056$\pm$0.039  \\
\hline                                         
CLEO~\cite{pipienucleo}       &  1.48$\pm$0.15$\pm$0.05  & 0.83$\pm$0.11$\pm$0.04 \\                         
\hline                               
\end{tabular}
\label{tab:com}
\end{center}
\end{table}

\begin{acknowledgments}
The BESIII Collaboration thanks the staff of BEPCII and the IHEP computing center for their strong support. This work is supported in part by National Key R\&D Program of China under Contracts Nos. 2023YFA1606000, 2023YFA1606704, 2020YFA0406300, 2020YFA0406400; National Natural Science Foundation of China (NSFC) under Contracts Nos. 12375092, 11635010, 11735014, 11935015, 11935016, 11935018, 12025502, 12035009, 12035013, 12061131003, 12192260, 12192261, 12192262, 12192263, 12192264, 12192265, 12221005, 12225509, 12235017, 12361141819; Guangdong Basic and Applied Basic Research Foundation under Grant No. 2023A1515010121 and the Fundamental Research Funds for the Central Universities under Contract No. 020400/531118010467; the Chinese Academy of Sciences (CAS) Large-Scale Scientific Facility Program; the CAS Center for Excellence in Particle Physics (CCEPP); Joint Large-Scale Scientific Facility Funds of the NSFC and CAS under Contract No. U1832207; 100 Talents Program of CAS; The Institute of Nuclear and Particle Physics (INPAC) and Shanghai Key Laboratory for Particle Physics and Cosmology; German Research Foundation DFG under Contracts Nos. 455635585, FOR5327, GRK 2149; Istituto Nazionale di Fisica Nucleare, Italy; Ministry of Development of Turkey under Contract No. DPT2006K-120470; National Research Foundation of Korea under Contract No. NRF-2022R1A2C1092335; National Science and Technology fund of Mongolia; National Science Research and Innovation Fund (NSRF) via the Program Management Unit for Human Resources \& Institutional Development, Research and Innovation of Thailand under Contract No. B16F640076; Polish National Science Centre under Contract No. 2019/35/O/ST2/02907; The Swedish Research Council; U. S. Department of Energy under Contract No. DE-FG02-05ER41374.
\end{acknowledgments}


\end{document}